\documentclass[a4paper,twocolumn,pra,showpacs,superscriptaddress]{revtex4}
\pdfoutput=1

\newcount\quick		% use png graphics for high-res plots
\usepackage{amsmath,amsfonts,amssymb}
\usepackage{graphicx}
\usepackage{enumerate}
\usepackage{bbm}
\usepackage{color}
\usepackage{epstopdf}

\newcommand{\ie}{\emph{i.e.}}

\newcommand{\si}[1]{\ensuremath{_{\text{#1}}}}
\newcommand{\vect}[1]{\ensuremath{\vec{#1}}}
\newcommand{\tens}[1]{\ensuremath{\boldsymbol{#1}}}
\newcommand{\T}{\ensuremath{^{\mathrm{T}}}}
\newcommand{\1}{\ensuremath{\mathbbm{1}}}

\newcommand{\ket}[1]{\ensuremath{|#1\rangle}}

\newcommand{\smp}{\Psi}

\newcommand{\pp}{V}

\newcommand{\thick}{\delta}

\newcommand{\scaledm}{m}

\newcommand{\wallp}[1]{\emph{#1}}

\newcommand{\eqna}{a}
\newcommand{\eqnb}{b}
\newcommand{\eqnA}{A}

\newcommand{\da}{u}
\newcommand{\db}{v}
\newcommand{\dc}{w}

\DeclareMathOperator{\real}{Re}

\begin{document}
 
\title{Optimized magnetic lattices for ultracold atomic ensembles}
\date{\today}

\author{Roman Schmied}
\email{Roman.Schmied@unibas.ch}
\affiliation{Max-Planck-Institut f\"ur Quantenoptik, Garching, Germany}
\affiliation{Departement Physik, Universit\"at Basel, Switzerland}
\author{Dietrich Leibfried}
\affiliation{National Institute of Standards and Technology, Boulder, Colorado, U.S.A.}
\author{Robert J. C. Spreeuw}
\affiliation{Van der Waals--Zeeman Instituut, Universiteit van Amsterdam, the Netherlands}
\author{Shannon Whitlock}
\email{S.M.Whitlock@uva.nl}
\affiliation{Van der Waals--Zeeman Instituut, Universiteit van Amsterdam, the Netherlands}

\pacs{37.10.Gh, 37.10.Jk, 41.20.Gz, 03.67.Lx}
% 03.67.Lx	Quantum computation architectures and implementations
% 37.10.Jk	Atoms in optical lattices
% 37.10.Gh	Atom traps and guides
% 41.20.Gz	Magnetostatics; magnetic shielding, magnetic induction, boundary-value problems

\begin{abstract}
We introduce a general method for designing tailored lattices of magnetic microtraps for ultracold atoms, on the basis of patterned permanently magnetized films. A fast numerical algorithm is used to automatically generate patterns which provide optimal atom confinement while respecting desired lattice symmetries and trap parameters. The algorithm can produce finite and infinite lattices of any plane symmetry; we focus specifically on square and triangular lattices which are of interest to future experiments. Typical trap parameters are discussed as well as the impact of realistic imperfections such as finite lithographic resolution and magnetic inhomogeneity. The presented designer lattices open new avenues for quantum simulation and quantum information processing with ultracold atoms on atom chips.
\end{abstract}
 
\maketitle

\section{Introduction}

Ultracold atoms in periodic potentials are ideal systems for exploring a diverse range of new physics, from the behavior of individual trapped atoms to rich many-body phenomena. At the core of most current experiments are optical lattices, produced by the interference of intersecting laser beams~\cite{Jaksch1998,Morsch2006,Bloch2008}, which have provided for example unprecedented access to Josephson effects~\cite{Albiez2005}, squeezed states~\cite{Esteve2008}, low-dimensional quantum gases~\cite{Paredes2004} and novel quantum phases~\cite{Greiner2002}. Particularly exciting applications arise in the context of quantum simulation~\cite{Lewenstein2007} and quantum information processing~\cite{DeMille2002,Treutlein2006b}, where atoms or molecules trapped in periodic arrays hold great promise for the storage and manipulation of quantum information. Lattices of magnetically trapped atoms or ensembles of atoms, interacting via a switchable Rydberg excitation, are promising systems for the generation of cluster states and other many-particle entangled states. Thus they could provide a suitable platform for implementing measurement-based quantum information processing schemes as well as exploring fault tolerant schemes~\cite{Raussendorf2001,Kitaev2003,Raussendorf2007,Weimer2010}. 

For practical devices, an attractive alternative to optical lattices are arrays of magnetic microtraps produced by current-carrying wires or patterned magnetic films on atom chips~\cite{Grabowski2003,Guenther2005,Sinclair2005,Ghanbari2006,Gerritsma2007,Boyd2007,Whitlock2009}. In principle, these magnetic lattices provide robust potentials for manipulating atoms, combined with a high degree of simplicity and excellent design flexibility. For example, inter-atomic distances are not restricted to certain fractions of optical wavelengths, allowing for a trade-off between inter-site coupling strength and single-site resolution. However, despite these advantages, only one-dimensional and a few two-dimensional lattices have been successfully implemented or proposed to date. The problem of finding the necessary magnetization pattern, or wire shape, which generates a desired set of atom traps currently relies on experience and trial-and error; what has been lacking is a general procedure for designing lattices with arbitrary arrangements of traps with the desired properties

In this paper we present a general method for creating tailored lattices of magnetic microtraps by controlling the geometric patterns of perpendicularly magnetized planar films. We employ a linear programming algorithm to find optimal single-layer magnetization patterns which produce desired lattice symmetries with specified trap parameters. This algorithm is similar to one previously used for the optimization of surface-electrode ion-trap lattices based on radio-frequency (rf) electric fields~\cite{Schmied2009}. It can be used to create a wide variety of designer-lattice geometries with arbitrary trap arrangements, opening new avenues for simulating condensed matter systems and for quantum information processing with ultracold atoms on atom chips. The planar patterns generated by the algorithm are simple to fabricate and can be straightforwardly implemented in existing experiments. Current-carrying wires tracing the edges of the optimized patterns represent an alternative experimental implementation of our results~\cite{Jackson}.

In the first part of the paper we give an overview of the defining properties of magnetic films (Section~\ref{sec:magnetization}) and Ioffe--Pritchard traps (Section~\ref{sec:IoffePritchard}), followed by a description of the linear programming algorithm (Section~\ref{sec:optimization}). We then turn to the results of the algorithm (Section~\ref{sec:optimised_lattices}). To highlight the flexibility of the method we focus on two desirable lattices we can generate: one square lattice and one triangular lattice, of interest for future experiments. These lattices offer tight confinement and a high degree of symmetry. In Section~\ref{sec:experimental} we consider typical trap parameters produced by these lattices and account for realistic imperfections such as finite lithographic resolution and inhomogeneity which could arise during fabrication. Section~\ref{sec:loading} discusses how these microtrap lattices can be loaded.

\section{Theoretical frame}

\subsection{Permanent magnetic microstructures}
\label{sec:magnetization}

For our present purposes of designing magnetic potentials we consider single-layer patterned magnetic thin films with perpendicular magnetization~\cite{Gerritsma2007,Whitlock2009}. There is an analogy between such patterned magnetic films and an effective boundary current which would produce an equivalent magnetic field~\cite{Jackson}.
We describe magnetic fields as gradients $\vect{B}(\vect{r})=-\vect{\nabla}\smp(\vect{r})$ of the \emph{scalar} magnetic potential $\smp(\vect{r})$~\cite{Jackson}. In particular, the scalar magnetic potential generated above a thin (typical thickness $\thick\ll z$) permanently out-of-plane magnetized layer of material in the $x y$ plane $\mathcal{P}$ is
\begin{multline}
	\label{eq:pot3D}
	\smp(x,y,z) =\\
	\frac12 \mu_0 \thick M_z
	\int_{\mathcal{P}} \scaledm(x',y') G(x-x',y-y',z)\text{d}x' \text{d}y',
\end{multline}
where $\mu_0$ is the permeability of free space and $M_z$ is the remanent out-of-plane magnetization. Typical magnetic films achieve permanent out-of-plane magnetizations of the order of $M_z\sim10^6$\,A/m, and typical film thicknesses are of the order of $\thick\sim100\,\text{nm}$~\cite{Gerritsma2007} (giving a magnetization current, or edge current in a wire implementation, of $\thick M_z\approx0.1\,\text{A}$); typical coercivities are around 500-1000\,G, such that any external fields applied to create the microtraps do not cause remagnetization. The dimensionless function $\scaledm(x,y)$ represents the spatial dependence of the magnetization current (\ie, the product of remanent magnetization and film thickness); we assume it to vary between $0$ and $1$ without loss of generality. The Green's function is
\begin{equation}
	\label{eq:Green}
	G(x,y,z) = \frac{z}{2\pi(x^2+y^2+z^2)^{3/2}}
\end{equation}
and serves to propagate the planar Dirichlet boundary condition $\smp(x,y,0)=\frac12 \mu_0 \thick M_z \scaledm(x,y)$ through space while satisfying Laplace's equation $\nabla^2\smp(\vect{r})=0$.

We restrict our analysis to single-layer magnetization patterns with binary ``step-like'' magnetization (or thickness) variations, as they are relevant for current experiments due to the ease of patterning~\cite{Gerritsma2007}. We note that the form of Eq.~\eqref{eq:pot3D} is proportional to that which propagates an \emph{electric} potential from an electrode plane into space~\cite{Schmied2010}. Further, the shape constraints on a magnetization pattern are mathematically equivalent to those on rf electrodes for ion trapping, as will be shown in Section~\ref{sec:IoffePritchard}. This leads us to the conclusion that the optimization algorithm of Ref.~\cite{Schmied2009}, recently used to generate electrode patterns for ion trapping, is also ideally suited to designing new geometries of magnetic lattices for ultracold neutral atoms. In this sense the main goal of the remainder of this article is to find specific magnetization patterns $\scaledm(x,y)$ which produce magnetic lattices of optimal strength (\ie, the stiffest and deepest magnetic traps for fixed magnetization current $\thick M_z$), given a set of constraints on trap arrangement and geometry.

Despite the mathematical analogies, there are several practical differences between the designs of rf ion traps and magnetic microtraps. These differences concern the spatial derivatives of the electric or magnetic scalar potentials necessary to produce stable trapping potentials. In what follows we detail the conditions necessary for generating magnetic trapping geometries of interest to current experiments.

\subsection{Ioffe--Pritchard traps}
\label{sec:IoffePritchard}

Magnetostatic traps confine ultracold atoms of a specific hyperfine state $\ket{F,m_F}$ through a spatially varying Zeeman shift pseudo-potential~\cite{Gerritsma2006}
\begin{equation}
	\label{eq:pseudopot}
	\pp(\vect{r}) = m_F g_F \mu\si{B} \|\vect{B}(\vect{r})\|,
\end{equation}
where $g_F$ is the level's Land\'e factor and $\mu\si{B}$ is the Bohr magneton. Unlike for rf ion traps~\cite{Schmied2009}, here the trapping minimum of $V(\vect{r})$ must be nonzero in order to prevent Majorana spin flips~\cite{Gerritsma2006}. Ioffe--Pritchard (IP) type traps are the simplest traps which guarantee this property.

Let us assume that we wish to trap atoms at several points $\vect{r}^{(\ell)}$ in space, with $\ell=1,2,3,\ldots$ indexing the various potential minima. Since the properties of each atom trap depend mostly on the local shape of the magnetic field, we require precise ways of specifying this shape in order to generate a desired lattice of atom traps. We base our analysis on a series expansion of the scalar magnetic potential around the trap points,
\begin{multline}
	\label{eq:series}
	\smp(\vect{r}) = \smp_0^{(\ell)} + \sum_{i=1}^3 \da_i^{(\ell)} (r_i-r_i^{(\ell)})\\
	+ \frac12 \sum_{i,j=1}^3 \db_{i,j}^{(\ell)} (r_i-r_i^{(\ell)})(r_j-r_j^{(\ell)})\\
	+ \frac16 \sum_{i,j,k=1}^3 \dc_{i,j,k}^{(\ell)} (r_i-r_i^{(\ell)})(r_j-r_j^{(\ell)})(r_k-r_k^{(\ell)})
	+ \ldots,
\end{multline}
where $(r_1,r_2,r_3) = (x,y,z)$ are the Cartesian coordinates. The vectors $\vect{\da}^{(\ell)}$ have components $\da_i^{(\ell)}$, while $\tens{\db}^{(\ell)}$ and $\tens{\dc}^{(\ell)}$ are tensors with components $\db_{i,j}^{(\ell)}$ and $\dc_{i,j,k}^{(\ell)}$, respectively. In this notation, $\vect{\da}^{(\ell)}=-\vect{B}(\vect{r}^{(\ell)})$ specifies the local magnetic field at the desired trap position (apart from external bias fields), while $\tens{\db}^{(\ell)}$ is the local gradient tensor and $\tens{\dc}^{(\ell)}$ specifies the local magnetic field curvatures. As described in Ref.~\cite{Gerritsma2006} the coefficients of Eq.~\eqref{eq:series} must be fully symmetric under permutation of indices and satisfy
\begin{subequations}
\begin{align}
	\label{eq:trace2}
	\db_{1,1}^{(\ell)}+\db_{2,2}^{(\ell)}+\db_{3,3}^{(\ell)} &=0\\
	\label{eq:trace3}
	\dc_{i,1,1}^{(\ell)}+\dc_{i,2,2}^{(\ell)}+\dc_{i,3,3}^{(\ell)} &=0 \quad \forall i\in\{1,2,3\}
\end{align}
\end{subequations}
since $\nabla^2\smp=-\vect{\nabla}\cdot\vect{B}=0$. We use these vectors and tensors $\vect{\da}^{(\ell)}$, $\tens{\db}^{(\ell)}$, and $\tens{\dc}^{(\ell)}$ as input parameters for our optimization algorithm (Section~\ref{sec:optimization}) in order to specify the shape and orientation of the magnetic field at the trap locations. In what follows we detail how these derivatives of $\smp$ are constrained to specific forms for representing a particular desired lattice of IP traps.

The defining characteristics of IP traps are that each gradient tensor $\tens{\db}^{(\ell)}$ has a zero eigenvalue associated with its `Ioffe axis', and that the total local magnetic field (`Ioffe field' $\vect{B}\si{I}$) is parallel to this axis: $\tens{\db}^{(\ell)}\cdot\vect{B}\si{I}=0$. We therefore fix the field gradient tensors $\tens{\db}^{(\ell)}$ at the trap locations [for concrete examples we use Eq.~\eqref{eq:rotbij}].

Although it is possible to produce self-biased magnetic microtraps which require no external fields~\cite{Fernholz2008}, in practice a homogeneous external bias field $\vect{B}_0$ is usually applied to produce the Ioffe field $\vect{B}\si{I}$ simultaneously at all trap sites. This dramatically simplifies loading the microtraps (see Section~\ref{sec:loading}) and ensures that the magnetic potential remains finite away from the film surface (allowing for deeper confinement).
To ensure all traps have equal depths for a given bias field $\vect{B}_0$, we typically constrain the first-derivative vectors $\vect{\da}^{(\ell)}$ to be equal, and align the Ioffe axes at all trap positions $\vect{r}^{(\ell)}$. However, there are no \emph{a priori} constraints on the actual values of these vectors and tensors.

If desired, additional constraints can be placed on the remaining two eigen-directions or the eigenvalues of the $\tens{\db}^{(\ell)}$, on the local fields $\vect{\da}^{(\ell)}$, or on the curvature tensors $\tens{\dc}^{(\ell)}$ in order to tailor the shapes, orientations and aspect ratios of the resulting trapping potentials. In particular, the curvature matrix of an IP trap, which defines the characteristic oscillation frequencies (trap stiffness), is given by
\begin{equation}
	\label{eq:ppcurv}
	\left. \frac{\partial^2\|\vect{B}(\vect{r})-\vect{B}(\vect{r}^{(\ell)})+\vect{B}\si{I}\|}{\partial r_i \partial r_j} \right|_{\ell}
	= \left[Ê\frac{\tens{\db}^{(\ell)}\cdot \tens{\db}^{(\ell)}+\tens{\dc}^{(\ell)}\cdot\vect{B}\si{I}}{\|\vect{B}\si{I}\|} \right]_{i,j}.
\end{equation}
It is important that all eigenvalues of these curvature matrices be positive for a certain homogeneous bias field $\vect{B}_0=-\vect{B}(\vect{r}^{(\ell)})+\vect{B}\si{I}$, in order to simultaneously trap atoms at all the positions $\vect{r}^{(\ell)}$. Unlike for rf ion traps~\cite{Schmied2009}, the three principal curvatures of IP traps are quite independent; prolate, spherical, oblate, and triaxial traps can be designed by judiciously choosing the conditions on the $\tens{\db}^{(\ell)}$ and $\tens{\dc}^{(\ell)}$ as well as operating at specific Ioffe field strengths.

All constraints on the derivatives of $\smp$ are only specified \emph{relative} to each other, not by their absolute values. The optimization algorithm of Section~\ref{sec:optimization} aims at discovering the magnetization pattern which satisfies the constraints outlined in this section while maximizing the absolute values of the derivatives of $\smp$. The fact that Eq.~\eqref{eq:ppcurv} depends on the Ioffe field strength makes it difficult to introduce a dimensionless quantity describing the absolute strength of an optimized IP trap. However, it appears that jointly maximizing the $\tens{\db}^{(\ell)}$ and $\tens{\dc}^{(\ell)}$ using our algorithm generally leads to deeper traps with higher trap frequencies, for fixed other physical constraints like the magnetization current $\thick M_z$.

\subsection{Optimization algorithm}
\label{sec:optimization}

The algorithm for finding the optimal magnetization pattern implementing a desired Ioffe--Pritchard trap array follows from Ref.~\cite{Schmied2009}. We present it here in a more extended form.

The first step in optimizing the magnetization pattern $\scaledm(x,y)$ consists of subdividing the atom chip surface into $N$ small domains. The shapes of these domains are irrelevant for our purposes, and their number only serves to increase the resolution of the final magnetization pattern. It is not necessary that these domains cover the entire atom chip, that they lie in a single plane, or that they form a simply connected subset of the plane. For finite sets of traps the optimization may be performed on a finite-sized pattern in order to fully account for boundary effects. For large periodic lattices of traps it is useful to divide a single unit cell of the wallpaper group with the desired lattice symmetry into domains, and then assume that the same pattern is repeated indefinitely over the atom chip surface.

We assume that within a domain $\alpha$ covering an area $\mathcal{P}_{\alpha}$ of the atom chip surface, the scaled magnetization $m(x,y)=\scaledm_{\alpha}$ is constant. The total scalar magnetic potential, Eq.~\eqref{eq:pot3D}, can then be decomposed into a sum over domains,
\begin{equation}
	\label{eq:smpdecomp}
	\smp(\vect{r}) = \frac12 \mu_0 \thick M_z \sum_{\alpha} \scaledm_{\alpha} \psi_{\alpha}(\vect{r}),
\end{equation}
where the functions
\begin{equation}
	\label{eq:pixelpot}
	\psi_{\alpha}(x,y,z)=\int_{\mathcal{P}_{\alpha}} G(x-x',y-y',z)\text{d}x'\text{d}y'
\end{equation}
are the scalar magnetic potentials induced by the domains indexed by $\alpha$, assuming unit magnetization (since the term $\frac12 \mu_0 \thick M_z \scaledm_{\alpha}$ has been factored out). 
The present algorithm can also be applied to different problems such as non-planar geometries, electrode patterns, or hybrid atom-ion traps by replacing Eq.~\eqref{eq:pot3D} with more general expressions, or by evaluating Eq.~\eqref{eq:pixelpot} numerically using finite-element or boundary-element methods~\cite{Wrobel}.
The derivatives of the scalar magnetic potential are similarly parametrized
\begin{equation}
	\label{eq:smpdecomp1}
	\frac{\partial\smp(\vect{r})}{\partial r_i} = \frac12 \mu_0 \thick M_z \sum_{\alpha} \scaledm_{\alpha} \frac{\partial \psi_{\alpha}(\vect{r})}{\partial r_i}
\end{equation}
etc. With these definitions the linear conditions of Section~\ref{sec:IoffePritchard} can all be expressed in terms of the scaled domain magnetizations $\scaledm_{\alpha}$ without knowing their values \emph{a priori}.  We define $\vect{\scaledm}$ as the vector with components $m_{\alpha}$, which represents the optimal magnetization pattern to be found, restricted to $\scaledm_{\alpha}=0$ or $\scaledm_{\alpha}=1$ for non-magnetic or fully magnetized domains, respectively. Using this notation, each linear condition of Section~\ref{sec:IoffePritchard} can be formulated in the form $\vect{\eqna}_k\cdot\vect{\scaledm}=C \eqnb_k$, where $C$ is a common prefactor in all conditions and $k$ indexes the different constraints. As in Ref.~\cite{Schmied2009} the conditions of Section~\ref{sec:IoffePritchard} are never imposed as absolute numbers but only relative to each other, while their absolute magnitudes (through the prefactor $C$) are jointly maximized by the algorithm in order to achieve maximal strength of the Zeeman pseudo-potential for a given magnetization current $\thick M_z$. All these conditions can be jointly expressed as a matrix equation
\begin{equation}
	\label{eq:lineq}
	\tens{\eqnA}\cdot\vect{\scaledm}=C\vect{\eqnb},
\end{equation}
with $|C|$ to be maximized.

In periodic arrays of microtraps the magnetization pattern is most naturally expressed in terms of its Fourier amplitudes $\hat{\scaledm}(k_x,k_y)=\mathcal{F}[\scaledm(x,y)]$. Instead of using the Green's function of Eq.~\eqref{eq:Green}, the propagation of the scalar magnetic potential away from the atom chip then follows from the exponential damping of each Fourier component,
\begin{multline}
	\smp(x,y,z) =\\ \frac12 \mu_0 \thick M_z 
	\real\left[ \sum_{k_x,k_y}Ê\hat{\scaledm}(k_x,k_y) e^{i(k_x x+k_y y)-z\sqrt{k_x^2+k_y^2}}\right],
	\label{eq:fouriermodes}
\end{multline}
where $\real[\ldots]$ refers to the real part.
Since the Fourier transform is a linear operation, these Fourier amplitudes $\hat{\scaledm}(k_x,k_y)$ can be expressed as linear functions of the domain magnetizations $\scaledm_{\alpha}$. Any linear constraints on the Fourier amplitudes can therefore also be brought into the form $\vect{\eqna}_k\cdot\vect{\scaledm}=\eqnb_k$ and thus included in Eq.~\eqref{eq:lineq}. However, we will not be using such direct Fourier constraints in our examples (Section~\ref{sec:optimised_lattices}).

The physical constraints describing the design for an atom trap lattice are thus formulated in the matrix equation of Eq.~\eqref{eq:lineq}. If the number of domains $N$ is large, then this linear system of equations is highly underdetermined. Which of the solutions should we choose? The inhomogeneous solution $\vect{\scaledm}_0=\tens{\eqnA}^+\cdot\vect{\eqnb}$, calculated from the pseudo-inverse $\tens{\eqnA}^+$, incorporates the main structure of the magnetization pattern we will generate. However $\vect{\scaledm}_0$ does not satisfy the constraint that $\scaledm_{\alpha}\in\{0,1\}$: its components are not normalized in any useful way, and they take on a continuum of values. To remedy this we decompose the target solution into $\vect{\scaledm}=C\vect{\scaledm}_0+\vect{\scaledm}'$ with $\tens{\eqnA}\cdot\vect{\scaledm}'=0$. Any vector of this form satisfies Eq.~\eqref{eq:lineq}; but it serves to illustrate that the solution $\vect{\scaledm}$ should consist of the maximum possible contribution in the inhomogeneous direction $\vect{\scaledm}_0$ (since we wish to maximize $|C|$) plus just the right amounts of homogeneous solutions (in $\vect{\scaledm}'$) in order to satisfy the binary constraints. Now we define $\tens{\tilde{\eqnA}}=\tens{\eqnA}\cdot\left(\1-\vect{\scaledm}_0\vect{\scaledm}_0\T/\|\vect{\scaledm}_0\|^2\right)=\tens{\eqnA}-\vect{\eqnb}\,\vect{\scaledm}_0\T/\|\vect{\scaledm}_0\|^2$, for which $\tens{\tilde{\eqnA}}\cdot\vect{\scaledm}=\tens{\eqnA}\cdot\vect{\scaledm}'=0$ because $\vect{\scaledm}'\cdot\vect{\scaledm}_0=([\tens{\eqnA}^+]^T\cdot\vect{\scaledm}')\cdot\vect{\eqnb}=0$. The optimization problem can then be expressed as a homogeneous \emph{binary integer linear program} $(\tens{\tilde{\eqnA}},\vect{\scaledm}_0,\{0,1\})$, which schematically reads thus:
\begin{enumerate}[(i)]
	\label{ref:linearprogram}
	\item \label{it:homogeneous} We seek a vector $\vect{\scaledm}$ within the space satisfying the linear constraints $\tens{\tilde{\eqnA}}\cdot\vect{\scaledm}=0$. This gives a set of homogeneous conditions on $\vect{\scaledm}$ which do not involve the constant $C$, and which express the fact that the solution $\vect{\scaledm}$ must satisfy the design constraints for the atom trap configuration, Eq.~\eqref{eq:lineq}.
	\item \label{it:binary}ÊAll components of the vector $\vect{\scaledm}$ must satisfy $\scaledm_{\alpha}\in\{0,1\}$. This condition expresses the desire for a binary magnetization pattern, consisting only of non-magnetic domains ($\scaledm_{\alpha}=0$) and fully magnetized domains ($\scaledm_{\alpha}=1$).
	\item We wish to maximize $|C|=|\vect{\scaledm}\cdot\vect{\scaledm}_0|/\|\vect{\scaledm}_0\|^2$, the ``strength'' of the solution in the inhomogeneous direction satisfying the design constraints. Larger values of $|C|$ yield stronger atom confinement for a given magnetization current $\thick M_z$.
\end{enumerate}
Such integer linear programs are known to be NP-hard to optimize~\cite{Papadimitriou}, if an exact solution even exists. Fortunately, and perhaps surprisingly, we can relax condition~(\ref{it:binary}) to $0\le\scaledm_{\alpha}\le 1$ $\forall \alpha=1\ldots N$ without loss of accuracy but drastically decreasing the computational complexity to $\mathcal{O}(N)$, making large numbers of domains (high resolution) feasible. After relaxing condition~(\ref{it:binary}) we are dealing with a \emph{linear program} $(\tens{\tilde{\eqnA}},\vect{\scaledm}_0,0\ldots 1)$, for which very efficient algorithms are available~\cite{MatousekGaertner}. The solution to the linear program is known to be globally optimal~\cite{MatousekGaertner}, and the obtained solutions consist of relatively large uniform patches of magnetization railed at either $\scaledm(x,y)=0$ or $\scaledm(x,y)=1$. Further it is known that the number of domains $\alpha$ which are \emph{not} railed is equal to the number of independent constraints in Eq.~\eqref{eq:lineq}~\cite{Schmied2009}. Since the number of domains can be made arbitrarily large, we are thus guaranteed to find solutions in which the ratio of un-railed to railed domains can be made vanishingly small. In practical cases the few un-railed domains usually come to lie at the edges between large railed patches and can be safely rounded to 0 or 1.

\subsubsection{limitations of the algorithm}
\label{sec:limits}

The optimization algorithm of the preceding section is a powerful new tool for designing magnetic lattice potentials with arbitrary geometries and trap characteristics.
There are, however, some limitations which deserve mention. 

Although any magnetization pattern generated by the above algorithm is guaranteed to produce the desired lattice of magnetic microtraps, it disregards the shape of the Zeeman pseudo-potential \emph{between} microtraps. While this point is not critical in ion trapping applications~\cite{Schmied2009}, it may be of great interest for atom trapping to have efficient tunneling paths between microtraps. To this end it may be necessary to add further linear constraints on the scalar magnetic potential.

Generally, for trap stability the resultant Zeeman pseudo-potentials should not contain any points of zero magnetic field in the vicinity of the traps, since they will lead to trap losses due to tunneling from the IP traps and subsequent Majorana spin flips~\cite{Gerritsma2006}. It is therefore more important than in ion trap setups that after a magnetization pattern $\scaledm(x,y)$ is optimized, we check the corresponding Zeeman pseudo-potential
\begin{equation}
	\label{eq:totalZpot}
	\pp(\vect{r}) = m_F g_F \mu\si{B} \|\vect{B}(\vect{r})-\vect{B}(\vect{r}^{(\ell)})+\vect{B}\si{I}\|
\end{equation}
for spurious null points. Such points can appear at unexpected locations which depend on the external bias field $\vect{B}_0=-\vect{B}(\vect{r}^{(\ell)})+\vect{B}\si{I}$, and can often be eliminated through additional constraints on $\smp$. To enforce complete elimination at the design stage of the algorithm would require non-local field constraints, which are difficult to implement directly in the optimization algorithm. Fortunately, for geometries involving only a few constraints (like the examples in Section~\ref{sec:optimised_lattices}), the optimized magnetic potentials rarely contain any field zeros, as spurious minima would reduce the achievable Zeeman pseudo-potential strength at the nominal trap positions. In more complex cases such as magnetic superlattices with multiple traps per unit cell, or lattices of traps involving higher curvature constraints [for example from Eq.~\eqref{eq:ppcurv}], it may become favorable for magnetic field zeros to occur at points of special symmetry or even anywhere in the trap lattice. In these situations extra care should be taken in choosing appropriate additional constraints for the algorithm to eliminate or shift the spurious zeros from the trapping regions. 

It may be of interest to directly constrain the symmetry of the Zeeman pseudo-potential. As an example,  the triangular lattice we present in Section~\ref{sec:tri2} has a high degree of symmetry with equal barriers between sites chosen to allow for symmetric tunneling. However, direct symmetry constraints on the potential are quadratic in the coefficients $\scaledm_{\alpha}$ and thus cannot be incorporated into the linear form of the algorithm. This could be partly overcome through the use of nonlinear optimization algorithms which allow both linear and nonlinear constraints on the magnetic field and pseudo-potential respectively. We have applied a nonlinear optimization algorithm to some problems; however, we find it considerably slower and limited in the achievable resolution when compared with the linear programming algorithm.

\subsubsection{implementation}

The linear optimization algorithm has been implemented independently in the Mathematica and Matlab programming environments for the general case of two-dimensional Bravais lattices, where each unit cell consists of $N=n_1\times n_2$ identical parallelogram-shaped domains. Further, a Mathematica implementation allows optimizing non-periodic (finite) patterns with arbitrary polygonal domain shapes. All implementations use primal-dual interior-point algorithms for solving the linear programs~\cite{Mehrotra1992} and yield consistent results for the obtained optimal magnetization patterns. The typical number of domains (or Fourier modes) used in the periodic optimization is $N\sim 200\times200$ which yields smooth magnetization patterns and is a sufficiently high resolution to ensure that the corresponding magnetic potentials are precisely defined.

\section{Optimized lattices}
\label{sec:optimised_lattices}

In this section we present results for two-dimensional periodic lattices with a single microtrap per unit cell.
As discussed in Section~\ref{sec:IoffePritchard}, the simplest IP atom traps are constructed by constraining the magnetic field gradient to be of the form
\begin{equation}
\label{eq:rotbij}
	\tens{\db}\propto R_{\phi,\theta,\psi}\cdot
		\begin{pmatrix}
			1 & 0 & 0\\
			0 & -1 & 0\\
			0 & 0 & 0
		\end{pmatrix}
		\cdot R_{\phi,\theta,\psi}\T
\end{equation}
at the desired trap locations,
where $R_{\phi,\theta,\psi}$ is a rotation matrix with three Euler angles specifying the orientation of the IP trap axis, $\vect{\nu}=\{\sin\theta\sin\psi,\sin\theta\cos\psi,\cos\theta\}$, and the orientation $\phi$ of the two perpendicular axes around this direction. The local field vector $\vect{\da}$ can be left unconstrained since we are free to apply any external homogeneous bias field $\vect{B}_0$ in order to null the local magnetic field at the traps. We have run the optimization procedure for a large set of Euler angles and computed the corresponding optimal lattice geometries. In practice we find that IP traps with in-plane Ioffe axis orientation ($\theta=\pi/2$) provide good confinement, in contrast to out-of-plane traps ($\theta=0$) which typically result in weak or no confinement perpendicular to the surface [in the absence of curvature constraints, Eq.~\eqref{eq:ppcurv}]. We stress however that the algorithm described in Section~\ref{sec:optimization} is general, and constraints on all derivatives of the scalar potential ($\da_i$, $\db_{i,j}$, $\dc_{i,j,k}$, and higher) can be applied to obtain any desired trap shape or orientation. In the following we give a general and two specific examples of in-plane ($\theta=\pi/2$) IP magnetic lattices, chosen for their desirable properties for future planned experiments with ultracold atoms.

\subsection{Two-wave lattices}
\label{sec:twowave}

An interesting class of IP trap arrays with in-plane Ioffe axis can be described by the three-dimensional scalar magnetic potential
\begin{multline}
	\label{eq:twowave_gen_smp}
	\smp_{\zeta,\psi}(\vect{r}) = \hat{\smp}\left[ \cos\alpha\sin\left(\frac{2\pi x}{d}-\frac{2\pi y}{d\tan\zeta}\right)
	\right.\\
	\left. + \sin\alpha\sin\left(\frac{2\pi y}{d\sin\zeta}\right) \right]e^{-\frac{2\pi z}{d\sin\zeta}}
\end{multline}
with $\alpha=\tan^{-1}\frac{\cos(\psi+\zeta)}{\cos\psi}$. It forms an oblique lattice of IP traps above the points $\vect{r}=n_1\{d,0\}+n_2\{d\cos\zeta,d\sin\zeta\}$ with $(n_1,n_2)\in\mathbb{Z}^2$, where $d$ is the lattice period. Their Ioffe axis is $\{\sin\psi,\cos\psi,0\}$, in accord with Eq.~\eqref{eq:rotbij} for $\theta=\pi/2$ and $\phi=\pi/4$. The trapping height $h$ is selected through the bias field, which must cancel the field
\begin{equation}
	\vect{B}(0,0,h) = \frac{\hat{\smp}}{d} e^{-\frac{2\pi h}{d\sin\zeta}} \times \frac{2\pi\cos\alpha}{\cos\psi}\{-\cos\psi,\sin\psi,0\}
\end{equation}
perpendicular to the Ioffe direction. What is special about these trap arrays is that there is a particular Ioffe field strength
\begin{equation}
	\label{eq:IoffeSymm}
	B\si{I} = \frac{\hat{\smp}}{d} e^{-\frac{2\pi h}{d\sin\zeta}} \times \frac{\pi\cos\alpha\sin(2\psi+\zeta)}{\cos^2\psi\cos(\psi+\zeta)}
\end{equation}
such that the Zeeman pseudo-potential~\eqref{eq:totalZpot} is invariant under exchange of the lattice axes (wallpaper group \wallp{cmm}), irrespective of $\psi$. In particular this gives rise to equal tunneling barriers in the two lattice directions. It is interesting to note that this \wallp{cmm} symmetry is present neither in the magnetization pattern $\scaledm(x,y)\propto \smp(x,y,0)$ nor in the bias or Ioffe fields. Therefore, using the optimization algorithm of Section~\ref{sec:optimization} to find \emph{binary} magnetization patterns which optimally produce scalar potentials like Eq.~\eqref{eq:twowave_gen_smp}, we can obtain experimentally relevant and non-trivial systems which are otherwise difficult to discover using manual methods. In what follows we perform such calculations to construct approximately \wallp{cmm}-symmetric square and triangular lattices of IP traps.

\subsubsection{square lattice}
\label{sec:square_twowave}

Square lattices are commonly produced using optical potentials but are a considerable challenge to implement with magnetic lattices, as the high degree of symmetry typically results in points of zero magnetic field strength which lead to Majorana spin flips~\cite{Gerritsma2006}. However, from Eq.~\eqref{eq:twowave_gen_smp} with $\zeta=\pi/2$ such square lattices of IP traps are readily generated with our algorithm. All values of $\psi$ except integer multiples of $\pi/4$ are allowed; angles $\psi$ close to $\pi/4+n\pi/2$, $n\in\mathbb{Z}$, give maximal trapping depth.

\begin{figure}
\ifdefined\quick
	\includegraphics[width=7cm]{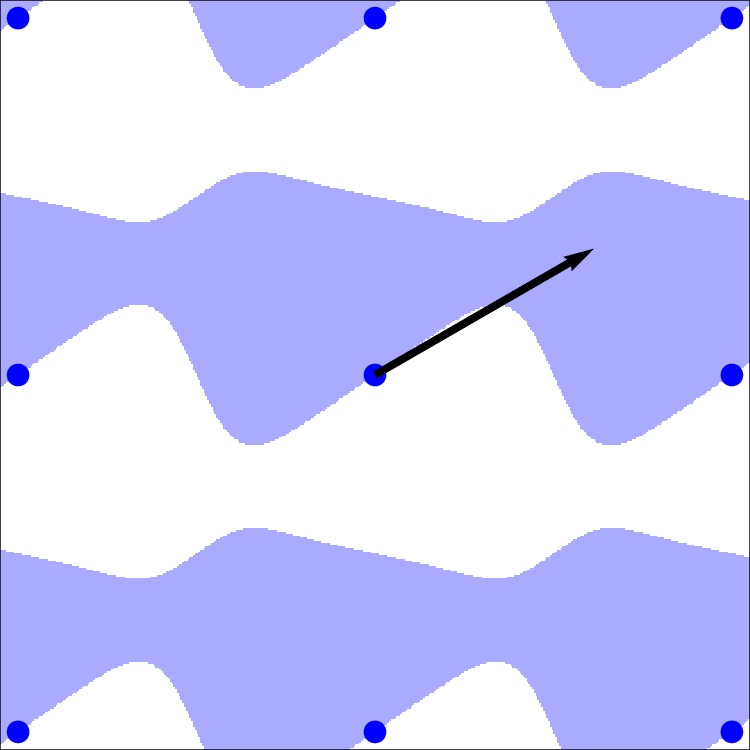}
\else
	\includegraphics[width=7cm]{square_pattern.pdf}
\fi
	\caption{Optimized magnetization pattern for a square lattice of IP traps (above blue points) constrained by Eq.~\eqref{eq:simple_curv}. Trapping height is $h=d/2$, and the Ioffe direction is given by the arrow ($\psi=\pi/3$). The optimization yields $C=1.06$ in Eq.~\eqref{eq:simple_curv}. Blue (white) areas are fully magnetized (unmagnetized) with $m=1$ ($m=0$).}
	\label{fig:square_simple_pattern}
\end{figure}

\begin{figure}
\ifdefined\quick
	\includegraphics[width=8.5cm]{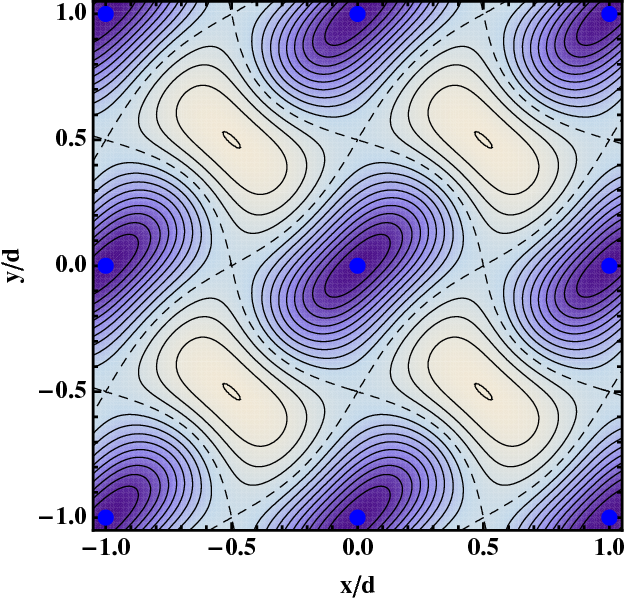}
\else
	\includegraphics[width=8.5cm]{square_potential.pdf}
\fi
	\caption{Zeeman pseudo-potential~[Eq.\eqref{eq:totalZpot}] in the trapping plane ($z=h=d/2$) produced by the magnetization pattern of Fig.~\ref{fig:square_simple_pattern} at a Ioffe field strength which creates equal potential barriers (dashed separatrix) between the potential wells (blue points) in the two lattice directions~\eqref{eq:IoffeSymm}. Darker colors signify deeper potentials.}
	\label{fig:square_simple_pot}
\end{figure}

A possible implementation of such a square lattice of IP traps is found by constraining the second derivative tensor of the scalar magnetic potential at the trapping sites to its value dictated by Eq.~\eqref{eq:twowave_gen_smp},
\begin{equation}
	\label{eq:simple_curv}
	\frac{\tens{\db}}{\frac12 \mu_0 \thick M_z/d^2} = C \times \begin{pmatrix}
				0 & 0 & \cos\psi\\
				0 & 0 & -\sin\psi\\
				\cos\psi & -\sin\psi & 0
			\end{pmatrix}.
\end{equation}
The first and third derivatives $\vect{\da}$ and $\tens{\dc}$ are left unconstrained. In Fig.~\ref{fig:square_simple_pattern} we show $2\times 2$ unit cells of the optimal magnetization pattern which generates this gradient tensor for $\psi=\pi/3$ and nominal trapping height $h=d/2$, \ie, half the inter-trap distance. The obtained pattern is typical in that it consist of large connected regions of uniform magnetization with equal areas of magnetized and non-magnetic regions. Furthermore, the pattern resembles the `staircase'-like structure recently experimentally demonstrated to create a two-dimensional lattice of IP traps~\cite{Gerritsma2007,Whitlock2009}. This shows that the optimization algorithm can produce magnetization patterns which resemble manually optimized lattice configurations.

The trapping height $h'$ is chosen by varying the bias field strength and direction. For $h'\ne h$ the curvature tensor may differ from Eq.~\eqref{eq:simple_curv}; however, $\det\tens{\db}(0,0,h')=0$ for any desired trap height, and therefore we are guaranteed to be able to produce IP traps everywhere on a vertical loading trajectory. Details of a method for loading such trap arrays are described in Section~\ref{sec:loading}.

Fig.~\ref{fig:square_simple_pot} shows the associated Zeeman pseudo-potential in the trapping plane, brought close to \wallp{cmm} symmetric by using an Ioffe field strength which creates equal potential barriers in the two lattice directions [similar to Eq.~\eqref{eq:IoffeSymm}]. Notice that the \wallp{cmm} symmetry is not exact because the higher-order Fourier modes used to construct the binary magnetization pattern (Fig.~\ref{fig:square_simple_pattern}) perturb the simple expression of Eq.~\eqref{eq:twowave_gen_smp}; neither the magnetization pattern nor the Ioffe direction are \wallp{cmm} symmetric, and therefore there is no \emph{a priori} reason why perfect \wallp{cmm} symmetry should be achievable in the Zeeman pseudo-potential. However, for sufficient nominal trapping heights $h/d$ this asymmetry may be negligible in practical applications (it is less than 3\% of the potential range in Fig.~\ref{fig:square_simple_pot}). Further, by a slight tuning of the Ioffe field strength we can force the \emph{effective} tunneling rates for trapped atoms in the two lattice directions to be exactly matched. This lattice is highly tunable: by varying the external bias field and the Ioffe field strength the barriers in the $x$ and $y$ lattice directions can be tuned independently, allowing time-dependent control over anisotropic tunneling rates.

\subsubsection{triangular lattice}
\label{sec:tri2}

\begin{figure}
\ifdefined\quick
	\includegraphics[width=7cm]{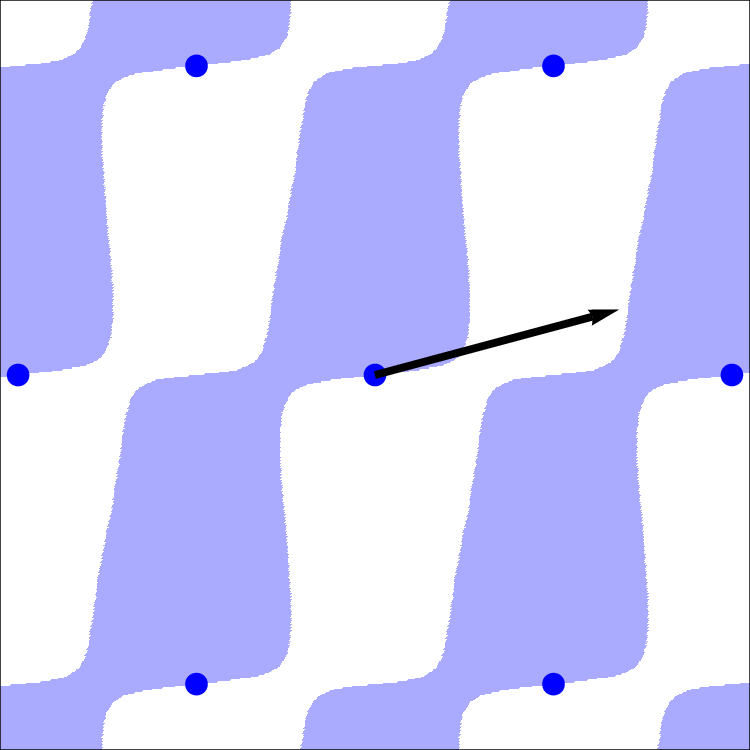}
\else
	\includegraphics[width=7cm]{triangular_pattern.pdf}
\fi
	\caption{Optimized magnetization pattern for a triangular lattice of IP traps (above blue points) constrained by Eq.~\eqref{eq:simple_curv} with $\psi=5\pi/12$ and additional field constraints at the potential barrier given in the text. The nominal trapping height is $h=d/2$, and the Ioffe direction is indicated by the arrow. The optimization yields $C=0.729$. Blue (white) areas are fully magnetized (unmagnetized) with $m=1$ ($m=0$).}
	\label{fig:triangular_simple_pattern}
\end{figure}

\begin{figure}
\ifdefined\quick
	\includegraphics[width=8.5cm]{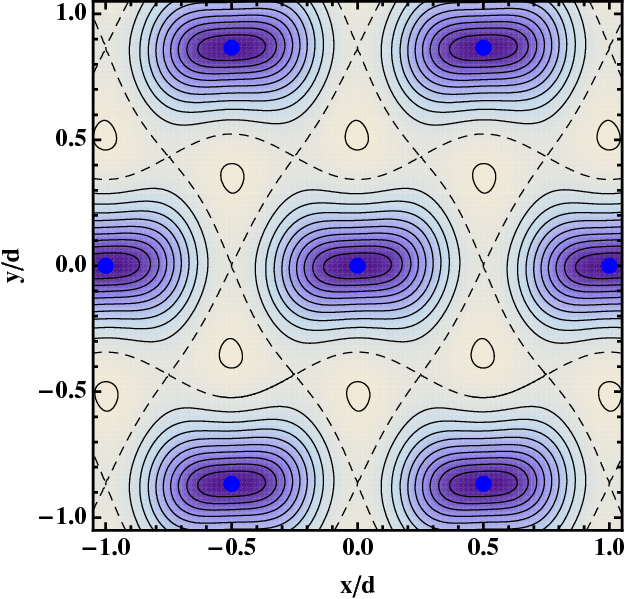}
\else
	\includegraphics[width=8.5cm]{triangular_potential.pdf}
\fi
	\caption{Zeeman pseudo-potential~[Eq.\eqref{eq:totalZpot}] in the trapping plane ($z=h=d/2$) produced by the magnetization pattern of Fig.~\ref{fig:triangular_simple_pattern} at a Ioffe field strength which creates equal potential barriers (dashed separatrix) between the potential wells (blue points) in the three lattice directions~\eqref{eq:IoffeSymm}. Darker colors signify deeper potentials.}
	\label{fig:triangular_simple_pot}
\end{figure}

Ultracold atoms in triangular lattices are of great experimental interest as they exhibit a wide variety of novel quantum phases~\cite{Schmied2008,Hauke2010,Eckardt2010}. However the considerable challenge involved in optically producing triangular lattices has so far limited experiments to a single recent realization~\cite{Becker2009}. The optimization algorithm is equally easily applied to produce triangular magnetic lattices, found from Eq.~\eqref{eq:twowave_gen_smp} with $\zeta=\pi/3$. For triangular lattices it would be desirable to have a Zeeman pseudo-potential which is not just \wallp{cmm}-symmetric, as given by Eq.~\eqref{eq:IoffeSymm}, but fully \wallp{p6m}-symmetric in order to have three equivalent tunneling directions and a truly isotropic lattice. Unfortunately, as mentioned in Section~\ref{sec:limits} such symmetry requirements involve nonlinear constraints beyond the linear programming method. However, we find that sufficiently symmetric lattices can be obtained by applying additional linear constraints at the barrier positions.

In Fig.~\ref{fig:triangular_simple_pattern} we present a magnetization pattern which produces a useful triangular lattice of microtraps with equal barrier heights in three directions, and nominal trapping height $h=d/2$. The required constraints include the gradient tensor at the trap sites given by Eq.~\eqref{eq:simple_curv} with $\psi=5\pi/12$, as well as an additional field constraint at one of the barrier positions at $\{x,y,z\}=\{d/2,0,d/2\}$. The required field constraint is
\begin{equation}
	\frac{\da_y}{\frac12 \mu_0 \thick M_z/d} = C\times (-0.0977),
\end{equation}
obtained automatically by nesting the linear programming algorithm within a nonlinear search algorithm aiming to null the pseudopotential barrier height difference (Fig.~\ref{fig:triangular_simple_pot}). This two-level approach is possible due to the speed of the linear programming algorithm. Although the resulting pseudopotential is not fully \wallp{p6m} symmetric, the effective tunnelling rates between traps could be matched experimentally by varying the Ioffe field strength $B\si{I}$ and the trap height to independently tune the potential barriers in all three directions.

As for the square lattice of the preceding section, the trapping height $h'$ is chosen by varying the bias field strength and direction, and a vertical loading trajectory of IP traps can be produced (see Section~\ref{sec:loading}).

\section{Experimental considerations}
\label{sec:experimental}

\subsection{Trapping parameters}

The lattice geometries of Section~\ref{sec:optimised_lattices} provide high trap depths and strong confinement for magnetically trapped atoms. In the following we calculate the resulting trap parameters, assuming realistic parameters for a FePt thin film~\cite{Gerritsma2007} with $M_z=670\,\text{kA}/\text{m}$ and $\delta=0.3\,\mu\text{m}$ (magnetization current $\thick M_z=0.2\,\text{A}$) trapping $^{87}$Rb atoms in the $F=2, m_F=2$ hyperfine state of the $5^2\text{S}_{1/2}$ electronic ground state ($g_1=-0.501\,826\,71(5)$ and $g_2=0.499\,836\,43(5)$ in this level~\cite{SteckRb87}).  A small but optically resolvable lattice period of $d=5\,\mu\text{m}$ is chosen so that nearest-neighbor interactions such as the Rydberg blockade effect would be observable~\cite{Urban2009,Gaetan2009}.

\begin{description}
\item[Square lattice] (Fig.~\ref{fig:square_simple_pattern}): A bias field of $B_0=\{-45.6,21.7,0\}$\,G creates traps at $h=2.5\,\mu\text{m}$ above the film layer, with an Ioffe field of $B\si{I}=-29.7\,\text{G}$. The barrier heights are equal in both $x$ and $y$ directions, $37.9\,\text{G}$ above the trap minimum. For $^{87}$Rb in the $\ket{2,2}$ state this corresponds to a barrier height of $2.55\,\text{mK}$. The trap depth away from the surface is $20.8\,\text{G}$ ($1.40\,\text{mK}$). Each trap is approximately cylindrically symmetric with long axis in the $\{1,1,0\}$ direction. The trap frequencies are $\omega_{z}=2\pi\times 124\,\text{kHz}$, $\omega_{\perp}=2\pi\times 121\,\text{kHz}$, and $\omega_{\parallel}=2\pi\times 37.6\,\text{kHz}$.
\item[Triangular lattice] (Fig.~\ref{fig:triangular_simple_pattern}): A bias field of $B_0=\{2.9,26.1,0\}\,\text{G}$ creates traps at $h=2.5\,\mu\text{m}$, with $B\si{I}=9.8\,\text{G}$. The barrier heights are equal in all three lattice directions, $34.6\,\text{G}$ ($2.32\,\text{mK}$) above the trap minimum. The trap depth away from the surface is $16.5\,\text{G}$ ($1.11\,\text{mK}$).  The trap frequencies are $\omega_{z}=2\pi\times 150\,\text{kHz}$, $\omega_{y}=2\pi\times 146\,\text{kHz}$, and $\omega_{x}=2\pi\times 36.9\,\text{kHz}$.
\end{description}
These lattices provide sufficiently tight confinement to localize the atoms to smaller than the optical wavelength. The Lamb--Dicke parameter is $\eta_j=\sqrt{\omega\si{recoil}/\omega_j}$ where $\omega\si{recoil}=2\pi\times3.771\,\text{kHz}$ for the $^{87}$Rb $\text{D}_2$ line. For the transverse dimensions $\eta<0.18$ and the atoms would enter the Lamb--Dicke regime ($\eta^2\langle n\rangle\ll 1$) for mean vibrational quanta $\langle n\rangle\leq30$. The magnetic fields and Zeeman pseudo-potential strengths for the given examples scale with $\thick M_z/d$ and the trap frequencies scale with $\sqrt{\thick M_z/d^3}$.

\subsection{Microfabrication}

The proposed magnetic lattices can be easily produced using conventional microfabrication techniques. Successful methods used in the past have included magneto-optical recording~\cite{Lau1999,Eriksson2004,Jaakkola2005}, hard-disk write head~\cite{Boyd2007}, grooved substrates and uniform film coating~\cite{Singh2009,Wang2005}, laser ablation~\cite{Wolff2009}, and optical or e-beam lithography followed by reactive ion etching~\cite{Xing2007,Gerritsma2007}. We favor lithography and ion etching since they provide greater freedom in the films used, arbitrary magnetization patterns are possible, and they can be readily extended to sub-micrometer resolution. Lithographic resolutions $\lesssim 50\,\text{nm}$ are possible with e-beam lithography, but long exposure times may be required to produce reasonably large lattices. Optical lithography is comparatively simple and can produce vast lattices; however the achievable resolution is limited by diffraction within the photo-resist layer to typically $1\,\mu\text{m}$. 

The optimized magnetic patterns are highly compatible with lithographic patterning, as they involve large connected magnetized regions with smooth boundaries. The effect of finite resolution is qualitatively equivalent to truncating the Fourier series expansion of the magnetic scalar potential. Because the contributions of higher Fourier modes (corresponding to small features) on the magnetic pseudo-potential decay rapidly with distance from the surface [Eq.~\eqref{eq:fouriermodes}], the shape of the potential at the trap position is relatively insensitive to the fine details of the magnetization pattern. For a more quantitative analysis, a perturbative study of the dependence of the pseudopotential on the magnetization in a given domain is readily performed with the parametrizations of Eqs.~\eqref{eq:smpdecomp}, \eqref{eq:smpdecomp1} etc.\ used in the optimization algorithm.

We have compared the trapping potentials produced by the optimized lattices at full numerical resolution to calculations with truncated Fourier series. Taking the triangular lattice as an example, we find that the lowest 13 Fourier modes ($\|\vect{k}\|\le 2\times\frac{2\pi}{d}$) are sufficient to adequately reproduce the desired potential (Fig.~\ref{fig:triangular_simple_pot}). This corresponds to an effective resolution of $\pi/\|\vect{k}\|=0.25d$. For the $d=5\,\mu\text{m}$ period lattice this would require a resolution of better than $\sim 1.2\,\mu\text{m}$, which can be achieved by optical lithography. With the reduced set of Fourier modes we find the pseudo-potential in the trapping plane maintains its symmetry and deviates from the full resolution calculation (Sec.~\ref{sec:tri2}) by at most 1.5\%. The trap depth increases by 0.4\% while the barrier heights decrease by 0.2-0.4\%; the trap frequencies increase by 0.4\%, 0.1\%, 1.1\% in the $z$, $y$, $x$ directions.

\subsection{Effect of inhomogeneity}

The quality of the magnetic lattices may also be affected by film roughness or magnetization inhomogeneity~\cite{Sinclair2005,Whitlock2007}. These lead to an additional randomly oriented magnetic field above the surface which causes small site-to-site variations in the potential experienced by the atoms. A simple model which takes into account random (white noise) fluctuations of the magnetization at a position above the edge of a magnetic film layer~\cite{Whitlock2007} is used to estimate the potential energy variations between lattice sites. A typical $rms$ surface roughness of around $5\,\text{nm}$ and grain size of $50\,\text{nm}$ would cause unwanted variations of less than $20\,\text{mG}$ corresponding to energy variations of less than $30\,\text{kHz}$. A more significant contribution may be due to imperfect magnetization and canting of magnetic domains. For a typical remanent-to-saturation magnetization ratio of 0.95 (squareness of the hysteresis loop) and a characteristic domain size of $50\,\text{nm}$, we anticipate additional variations of $60\,\text{mG}$ ($85\,\text{kHz}$). The estimated combined effect of surface roughness and magnetic inhomogeneity is comparable to or less than the vibrational level spacing and is therefore not expected to be a serious obstacle for experiments. The anticipated site-to-site variations could be reduced further with high quality Fe/Pt or Co/Pt multilayer thin-films, which can be atomically flat with small magnetic domains and higher squareness ratios~\cite{Lin1991}.

\subsection{Loading trajectory}
\label{sec:loading}

Atoms can be transferred to the proposed lattices simply using an auxillary macroscopic Z-shaped wire beneath the chip surface~\cite{Gerritsma2007}. An electric current through the Z-wire combined with an external bias field produces a single Ioffe--Pritchard magnetic trap~\cite{Reichel2002}, hundreds of micrometers above the film surface where the influence of the magnetic lattice is negligible. 

We have computed a loading trajectory which smoothly transfers atoms from the Z-wire trap to the magnetic lattice potential generated by the pattern of Fig.~\ref{fig:triangular_simple_pattern} and a lattice spacing of $d=5\,\mu\text{m}$. We assume a Z-wire positioned $0.4\,\text{mm}$ beneath the magnetic film, with a central segment of $L=1\,\text{mm}$ oriented such that its Ioffe axis matches that of the microtrap lattice at the nominal trap height $h=2.5\,\mu\text{m}$. In this symmetric situation $\det\tens{\db}(0,0,z)=0$ $\forall z>0$ for the total scalar magnetic potential (lattice + Z-wire), and thus for any desired trapping height $h'$ we can still find a homogeneous in-plane bias field $\vect{B}_0$ which creates IP traps~\cite{Gerritsma2006}. By smoothly ramping down the Z-wire current or increasing the bias field strength, the trap minimum moves toward the surface and the Z-wire trap eventually merges with the lattice microtraps.

\begin{figure}
	\hspace{-12pt}
	\ifdefined\redoepstopdf
		\includegraphics[width=8.25cm]{figurea.eps}
		\includegraphics[width=8.0cm,trim=0pt 40pt 0pt 0pt, clip=true]{figureb.eps}
		\includegraphics[width=8.0cm,trim=0pt 40pt 0pt 10pt, clip=true]{figurec.eps}
		\includegraphics[width=8.0cm,trim=0pt 0pt 0pt 10pt, clip=true]{figured.eps}
	\else
		\includegraphics[width=8.25cm]{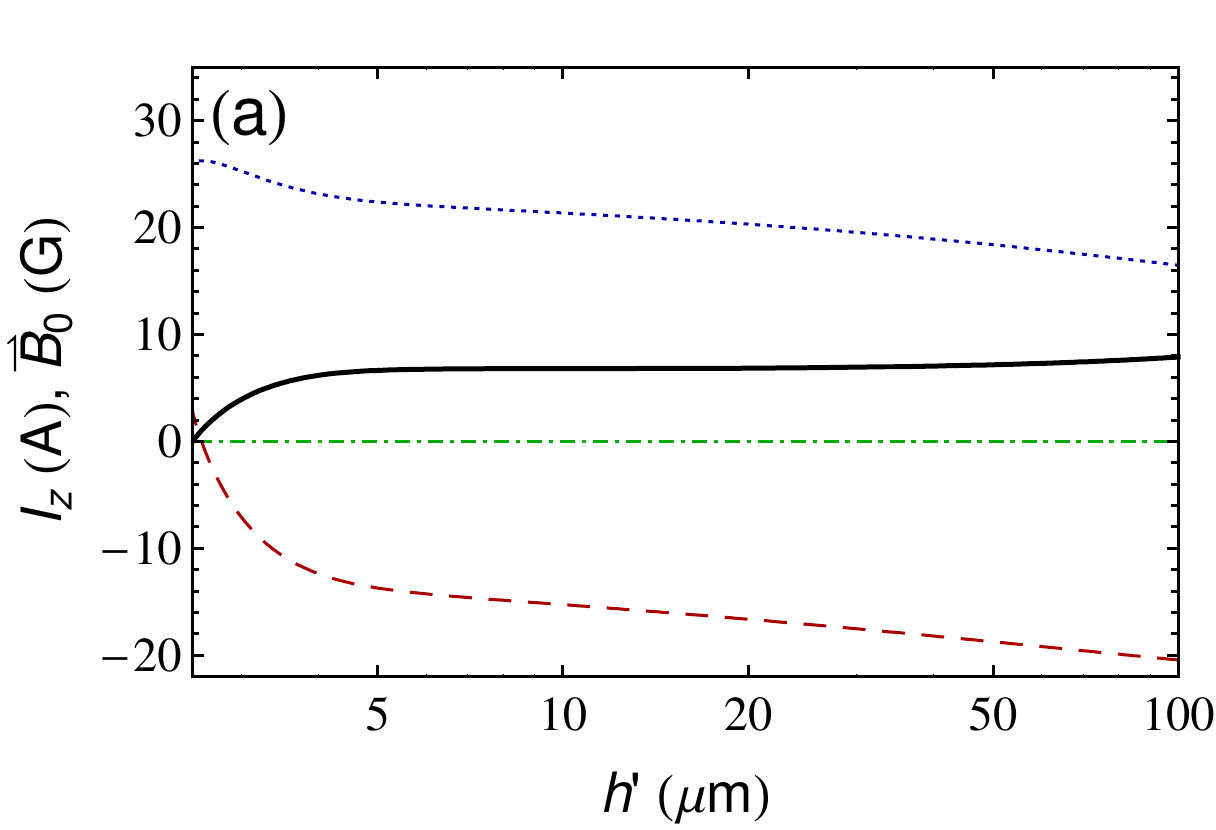}
		\includegraphics[width=8.0cm,trim=0pt 40pt 0pt 0pt, clip=true]{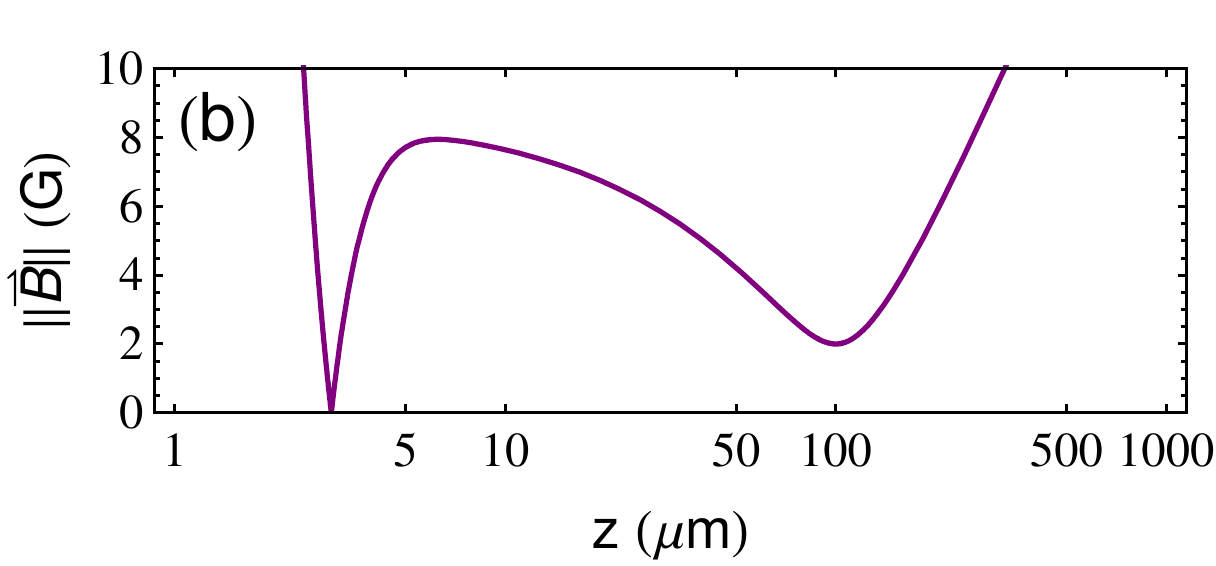}
		\includegraphics[width=8.0cm,trim=0pt 40pt 0pt 10pt, clip=true]{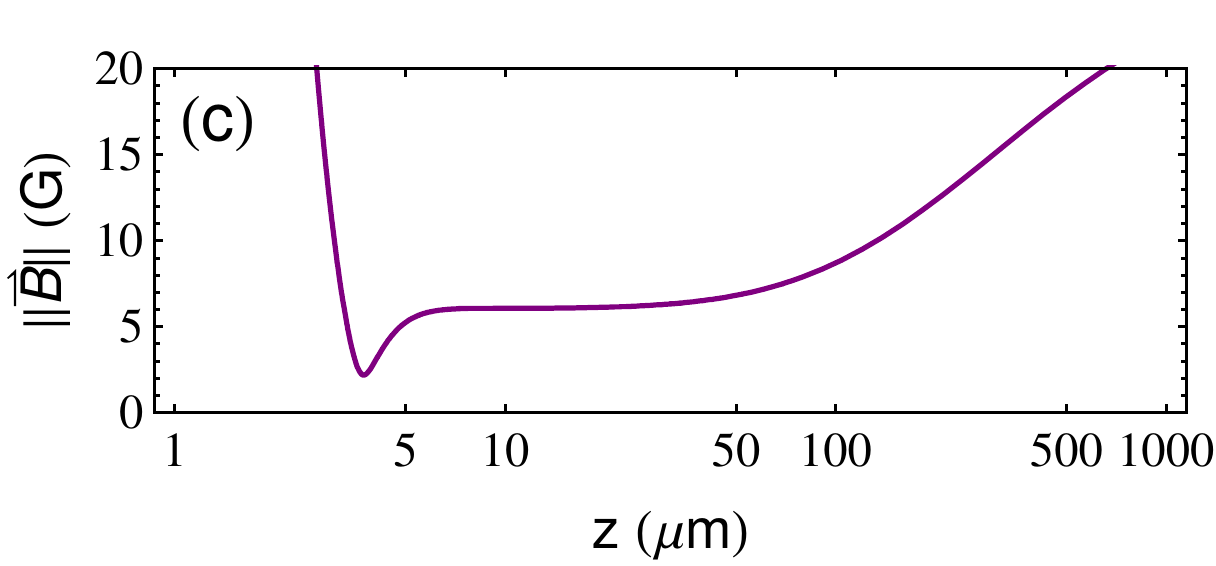}
		\includegraphics[width=8.0cm,trim=0pt 0pt 0pt 10pt, clip=true]{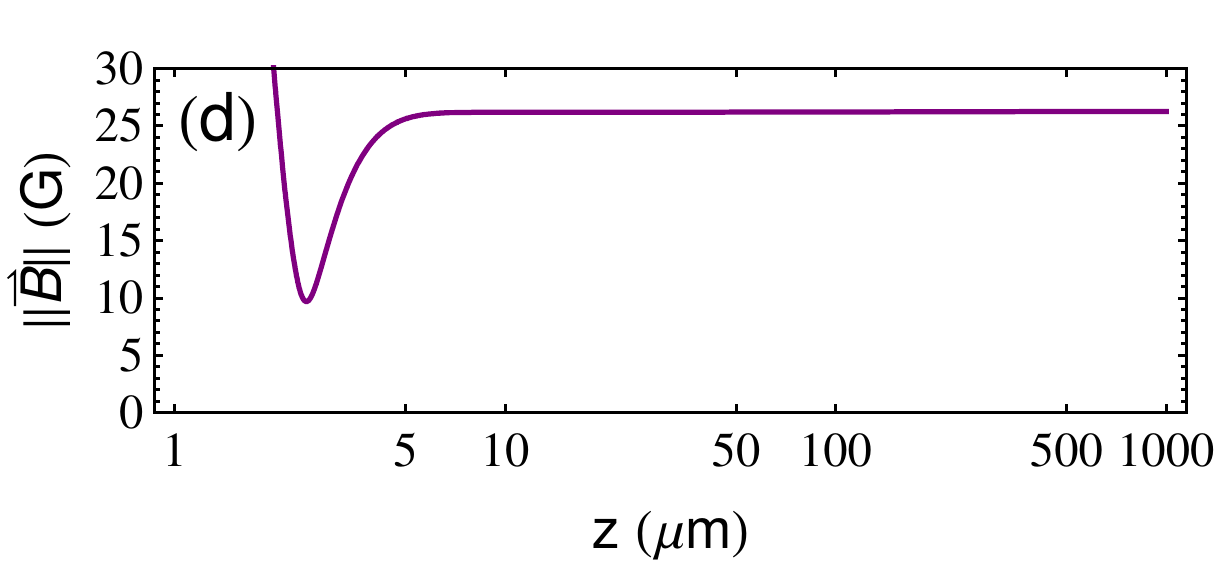}
	\fi
	\caption{Calculated trajectory for loading atoms to the triangular lattice. (a) Z-wire current $I_z$ (solid line) and applied bias field $\vect{B}_0$ (dashed, dotted, and dash-dotted lines for the $x$, $y$, and $z$ components, respectively) as functions of trapping height $h'$. Magnetic field strength $\|\vect{B}\|$ as a function of distance from the film surface: (b) prior to transfer ($h'=100\,\mu\text{m}$), (c) during transfer ($h'=10\,\mu\text{m}$) and (d) after transfer to the magnetic lattice is complete ($h'=h=2.5\,\mu\text{m}$). Note the logarithmic $z$ scale.}
	\label{fig:loading}
\end{figure}

For each trapping height between $h'=100\,\mu\text{m}$ and $h'=h=2.5\,\mu\text{m}$, the bias field ($\vect{B}_0$) and Z-wire current ($I_z$) are optimized to maintain a trap depth equal to or greater than the final depth in the lattice of $>16.5\,\text{G}$ to prevent loss of atoms during loading [Fig.~\ref{fig:loading}(a)]. The Ioffe-field strength is chosen to start at $B\si{I}=2\,\text{G}$ and increases during the trajectory to its final value of $B\si{I}=9.8\,\text{G}$ to create a symmetric triangular lattice (Fig.~\ref{fig:triangular_simple_pot}). Initially the atoms are confined in the large potential well of the Z-wire trap at $h'=100\,\mu\text{m}$, however the superimposed field of the lattice produces shallow secondary potential wells at $z\approx 3\,\mu\text{m}$ [Fig.~\ref{fig:loading}(b)]. The magnetic field strength as a function of height $z$ during the loading trajectory is calculated at the position of the global minimum in the $xy$ plane. As the fields are varied to move the Z-wire trap closer to the surface the minima approach, and at a distance of around $h'\approx 10\,\mu\text{m}$ atoms can spill over to the secondary traps [Fig.~\ref{fig:loading}(c)]. At this stage a thermal atom cloud would begin to fragment. Evaporative cooling in the combined Z-wire+lattice trap may be very efficient as the local phase-space density could be high in the small potential wells~\cite{Pinkse1997,StamperKurn1998,Ma2004}. Finally the Z-wire current is ramped off completely and the atoms are confined by the lattice potential alone [Fig.~\ref{fig:loading}(d)]. The precise number of traps that can be loaded in this way is difficult to estimate since it depends on the temperature of the clouds during loading and the rate of transfer. In practice, it seems possible to load a $1.5\times0.5\,\text{mm}^2$ region of the lattice~\cite{Whitlock2009}, corresponding to $\sim 3\times10^4$ populated microtraps. Throughout the transfer we ensure no field zeros are produced in the vicinity of the atoms, such that Majorana spin flip loss can be neglected. Additionally, the field strength at the film surface is $\|\vect{B}\|>120\,\text{G}$ in the $z=1\,\mu\text{m}$ plane, corresponding to a repulsive atom-surface barrier much greater than the attractive Casimir--Polder potential, thus preventing atom loss to the surface.

\section{Conclusions}

We have introduced a linear programming algorithm tailored to the problem of designing two-dimensional magnetic lattices of Ioffe--Pritchard traps for ultracold atoms. Previously, the design of such lattices required substantial experience and trial-and-error. The algorithm automatically generates single-layer binary magnetization patterns which produce desired lattice symmetries with specified trap parameters. It allows for designing non-trivial geometries which would be extremely difficult to obtain using manual methods.

The strengths of the algorithm were exemplified in a desirable square magnetic lattice and in the new case of a triangular magnetic lattice potential. The generated magnetic patterns consist of smooth connected regions of uniform magnetization, which are easily fabricated using existing methods. With realistic magnetic-film parameters these lattices will produce vast arrays of microscopic IP traps with tight confinement and high trap depths. The resulting potentials are of great interest for studies of quantum gases in lattice potentials~\cite{Greiner2002,Paredes2004,Becker2009}, condensed-matter analog systems~\cite{Lewenstein2007,Schmied2008,Hauke2010,Eckardt2010}, or quantum information processing with neutral atoms~\cite{DeMille2002,Treutlein2006b,Raussendorf2001,Kitaev2003,Raussendorf2007,Weimer2010}.

The similarities between the algorithms for designing magnetic Ioffe--Pritchard traps and electric radio-frequency (rf) ion traps could be exploited to produce hybrid traps, as a natural route to confine individually controlled rf-trapped ions within magnetically trapped ultracold quantum gases~\cite{Zipkes2010a,Zipkes2010b}. By appropriate changes of the external bias field the neutral atoms could then be moved to overlap with the ions or separated from them in a well-controlled manner. Further, once the ions are cooled to their motional ground state the rf traps can be switched off, transferring the ions to the magnetic microtraps (provided that their magnetic moments are appropriate) in order to eliminate rf micromotion~\cite{Berkeland1998,Schneider2010}. In practice one possible approach would be to add extra constraints in Eq.~\eqref{eq:lineq} to produce patterns that simultaneously provide magnetic traps for neutral atoms and, by applying rf voltages directly to the proposed magnetic structures (provided their rf impedance is suitable), rf traps for ions at the same sites or at sites of our choice.
Alternatively one can pattern an appropriately shaped conductive rf layer on top of the magnetic structures, thus decoupling the patterns for atom and ion traps.

\acknowledgments

We thank C.~F.~Ockeloen for initiating the calculations on loading the lattice microtraps.
R.~Schmied  was supported by the Deutsche Forschungsgemeinschaft (Forschungsgruppe 635) and the European Union (ACUTE). D.~Leibfried was supported by DARPA, NSA, ONR, IARPA, Sandia National Labs, and the NIST Quantum Information Program. S.~Whitlock acknowledges support from a Marie-Curie fellowship (PIIF-GA-2008-220794) and from the Stichting voor Fundamenteel Onderzoek der Materie (FOM), which is financially supported by the Nederlandse Organisatie voor Wetenschappelijk Onderzoek (NWO).

\bibliography{MPQ}

\end{document}